\documentclass{iopart}
\usepackage{graphicx}  
\usepackage{latexsym}
\usepackage{amssymb}

\jl{6}        
\eqnobysec    

\def\beq{\begin{equation}}
\def\eeq{\end{equation}}
\def\rmd{{\rm d}} 
\def\rmD{{\rm D}}

\begin{document}

\title[Spin precession along circular orbits in the Kerr spacetime]
{Spin precession along circular orbits in the Kerr spacetime: the Frenet-Serret description}

\author{
Donato Bini$^* {}^\S{}^\dag$,
Fernando de Felice$^\P$,
Andrea Geralico${}^\S$
and Robert T Jantzen$^\diamond {}^\S$
}
\address{
  ${}^*$\
Istituto per le Applicazioni del Calcolo ``M. Picone'', CNR I-00161 Rome, Italy
}
\address{
  ${}^\S$\
  ICRA,
  University of Rome, I-00185 Rome, Italy
}
\address{
  ${}^\dag$\
  INFN sezione di Firenze, I-00185
  Sesto Fiorentino (FI), Italy
}

\address{
  ${}^\P$\
  Dipartimento di Fisica, Universit\`a di Padova,  I-35131 Padova, Italy
}

\address{${}^\diamond$
Department of Mathematical Sciences, Villanova University, Villanova, PA 19085, USA
}

\begin{abstract}
The circular motion of spinning massive test particles in the equatorial plane of a rotating black hole is investigated in the 
case in which the components of the spin tensor are allowed to vary along the orbit. 
\end{abstract}

\pacno{04.20.Cv}

\section{Introduction}

The  equations of motion  for a spinning test particle in a given gravitational background were deduced
by Mathisson and Papapetrou \cite{math37,papa51}.
Letting $U$ be the timelike unit tangent vector of the ``center of mass line'' used to make the multipole reduction,
these equations are
\begin{eqnarray}
\label{papcoreqs1}
\frac{DP^{\mu}}{\rmd \tau_U}&=&-\frac12R^{\mu}{}_{\nu\alpha\beta}U^{\nu}S^{\alpha\beta}\equiv F^{\rm (sc)}{}^{\mu}\ , \\
\label{papcoreqs2}
\frac{DS^{\mu\nu}}{\rmd \tau_U}&=&P^{\mu}U^{\nu}-P^{\nu}U^{\mu}\ ,
\end{eqnarray}
where $P^{\mu}$ is the total four-momentum of the particle and $S^{\mu\nu}$ is a (antisymmetric) spin tensor, both fields defined only along the center of mass world line, for which $\tau_U$ is a proper time parameter. 
These equations evolve $P$ and $S$ along this world line.
By contracting both sides of Eq.~(\ref{papcoreqs2}) with $U_\nu$, one obtains the following expression for the total 4-momentum
\begin{equation}
\label{SchwPs}
P^{\mu}=-(U\cdot P)U^\mu -U_\nu \frac{DS^{\mu\nu}}{\rmd \tau_U}\equiv
mU^\mu +P_s^\mu\ ,
\end{equation}
where $m=-U\cdot P$ reduces to the ordinary mass in the case in which the particle is not spinning, and  
$P_s$ is a vector orthogonal to $U$.

The assumption that the particle under consideration is a test particle means that both its mass as well as its spin must be small enough not to contribute significantly to the background metric. 
Moreover, in order to have a closed set of equations, the above equations of motion must be completed 
with supplementary conditions (SC), for which the standard choice is one of the following:
\begin{itemize}
\item[1.]
Corinaldesi-Papapetrou \cite{cori51} conditions: $S^{\mu\nu}(e_0)_\nu=0$, where $e_0$ is the coordinate timelike direction given by the background,
\item[2.]
Pirani \cite{pir56} conditions: $S^{\mu\nu}U_\nu=0$, 
\item[3.]
Tulczyjew \cite{tulc59} conditions: $S^{\mu\nu}P_\nu=0$.
\end{itemize}
All of these are algebraic conditions on the spin tensor components. 

The Mathisson-Papapetrou model does not give {\it a priori}
restrictions on the causal character of $U$ and $P$ and there is no
agreement in the literature on how this point should be considered \cite{tod77,masspin2}.
When both $U$ and $P$ are timelike vectors like $e_0$, each of the above SCs simply state that the spin tensor is purely spatial with respect to the associated direction in time, or equivalently that the spin vector defined by duality lies in the corresponding local rest space.
Only solutions of the combined equations for which both $U$ and $P$ are timelike vectors are considered in the present paper in order to have a meaningful interpretation describing a spinning test particle with nonzero rest mass and physical momentum; furthermore, only the Pirani and Tulczyjew supplementary conditions are discussed, ignoring the unnatural Corinaldesi-Papapetrou condition.

We study the behavior of nonzero rest mass spinning test particles moving along circular orbits in the Kerr spacetime in the 
case in which the components of the spin tensor are allowed to vary along the orbit, generalizing some previous work \cite{bdfg2,bdfgj}. 
Our present analysis makes use of the invariant spacetime Frenet-Serret frame approach, which is especially useful for the description of this family of orbits.  
The Frenet-Serret formalism reflects only the geometrical
properties of spacetime and of the worldline, being defined in an invariant
manner without reference to any particular coordinate system or observer.
Iyer and Vishveshwara \cite{iyer-vish} have introduced the rather involved computational formulas
necessary for explicit evaluation of the Frenet-Serret frame $\{E_{\alpha}\}$, $\alpha=0,\ldots,3$ and its associated curvature $\kappa$
and torsions $\tau_1, \tau_2$ \cite{honig,synge} for arbitrary constant speed circular orbits in general
stationary axisymmetric spacetimes.

The problem is thus addressed within a more elegant framework with respect to the treatment developed in the companion paper \cite{bdfgj} concerning the special case of a Schwarzschild black hole. As a consequence, even though details are naturally more involved, results are actually more meaningful, all quantities being expressed in terms of the Frenet-Serret curvature and torsions.
The restriction to circular motion is a very strong limitation for the system. In fact, the effect of time varying spin on the acceleration of the test particle path will break the symmetry of that path unless the spin precession is very closely tied to the natural Frenet-Serret rotational properties of the path itself. We find that as in the nonrotating case, only the Pirani supplementary conditions permit such specialized solutions in the rotating case of a Kerr black hole since they allow the spin tensor to be described completely by a spatial vector with respect to $U$. 
Thus the spacetime rotation does not substantially change the features of the circular motion of spinning particles compared to the static case.
Abandoning the restriction to circular motion leads in general to non-periodic motion, a feature that seems to characterize the  general situation in the Schwarzschild \cite{maeda} as well as Kerr \cite{semerak,hartl1,hartl2} spacetimes.

In what follows indices either indicated by latin letters or explicit numbers refer to frame components.

\section{Equations of motion in the Frenet-Serret formalism}

The Frenet-Serret frame $\{E_{\alpha}\}$, $\alpha=0,\ldots,3$ (with dual frame $\{\Omega^\alpha\}$)
along a single timelike worldline with tangent 4-vector $U=E_0$ and parametrized by the proper time $\tau_U$ is described by the following system of evolution equations \cite{iyer-vish}
\begin{eqnarray}
\label{FSeqs}
\frac{DE_0}{d\tau_U}&=\kappa E_1\ , \qquad &  
\frac{DE_1}{d\tau_U}=\kappa E_0+\tau_1 E_2\ ,\nonumber \\
 \nonumber \\
\frac{DE_2}{d\tau_U}&=-\tau_1E_1+\tau_2E_3\ , \qquad &
\frac{DE_3}{d\tau_U}=-\tau_2E_2\ .
\end{eqnarray}

The absolute value of the curvature $\kappa$ is the magnitude of
the acceleration $a(U)\equiv DU/d\tau_U=\kappa E_1$, while
the first
and second torsions $\tau_1$ and $\tau_2$ are the components of the Frenet-Serret angular velocity vector
\beq
\label{omegaFS}
\omega_{\rm (FS)}=\tau_1 E_3 + \tau_2 E_1\ , \qquad ||\omega_{\rm (FS)}||=[\tau_1^2 + \tau_2^2]^{1/2}\ ,
\eeq
with which the spatial Frenet-Serret frame $\{E_a\}$ rotates with respect to a Fermi-Walker transported frame along $U$. In terms of $\omega_{\rm (FS)}$, the evolution equations (\ref{FSeqs}) for the spatial frame vectors can be written in the more compact form 
\beq
\label{compact}
\frac{DE_a}{d\tau_U}=\omega_{\rm (FS)}\times E_a+\kappa E_0\,\delta^1_a\ ,
\eeq
where 
\beq
\omega_{\rm (FS)}\times E_a=(\tau_1\epsilon_{3ab}+\tau_2\epsilon_{1ab})E_b\ ,
\eeq
$\epsilon_{ijk}$ ($\epsilon_{123}=1$) denoting the Levi-Civita alternating symbol.

Consider first the spin evolution equations (\ref{papcoreqs2}); their frame components are explicitly
\begin{eqnarray}\fl\quad
\label{FSspineqsgen}
\frac{\rmd S_{01}}{\rmd \tau_U}&=\tau_1S_{02}+P_s{}_1\ , \qquad &
\frac{\rmd S_{02}}{\rmd \tau_U}=-\tau_1S_{01}+\tau_2S_{03}+\kappa S_{12}+P_s{}_2\ , \nonumber\\
\fl\quad
\frac{\rmd S_{03}}{\rmd \tau_U}&=-\tau_2S_{02}+\kappa S_{13}+P_s{}_3\ , \qquad &
\frac{\rmd S_{12}}{\rmd \tau_U}=\kappa S_{02}+\tau_2 S_{13}\ , \nonumber\\
\fl\quad
\frac{\rmd S_{13}}{\rmd \tau_U}&=-\tau_2 S_{12}+\tau_1 S_{23}+\kappa S_{03}\ , \qquad &
\frac{\rmd S_{23}}{\rmd \tau_U}=-\tau_1 S_{13}\ .
\end{eqnarray}
The orthogonal decomposition (\ref{SchwPs}) of the total 4-momentum $P$ can be further refined by introducing a unit vector $U_s$ along its spatial part with respect to $U$
\beq
\label{PdefFS}
P=m U + P_s\equiv mU +m_s U_s\ .
\eeq
In terms of frame components, $P_s^a=(\rmD S/\rmd\tau_U)^{0a}$ and
$m_s=U_s{}_a (\rmD S/\rmd\tau_U)^{0a}$.

Consider now the momentum evolution equation (\ref{papcoreqs1}).
From Eqs.~(\ref{PdefFS}) and (\ref{FSeqs}), its left hand side can be written
\beq
\label{FSdPdtaugen}
\frac{DP}{\rmd \tau_U}=\frac{\rmd m}{\rmd \tau_U}E_0+m\kappa E_1+\frac{DP_s}{\rmd \tau_U}\ ,
\eeq
where Eq.~(\ref{compact}) yields
\beq
\label{FSdPsdtaugen}
\frac{DP_s}{\rmd \tau_U}=\frac{\rmd P_s^a}{\rmd \tau_U}E_a+\omega_{\rm (FS)}\times P_s+\kappa E_0P_s^1\ .
\eeq
The momentum equations of motion are therefore
\begin{eqnarray}\fl\quad
\label{FSeqsmotogen}
\qquad 0&=\frac{\rmd m}{\rmd \tau_U}+\kappa P_s^1\ , \qquad &
\frac{\rmd P_s^1}{\rmd \tau_U} =F^{\rm (sc)}{}^1-m\kappa+\tau_1 P_s^2\ , \nonumber\\
\fl\quad
\frac{\rmd P_s^2}{\rmd \tau_U} &=F^{\rm (sc)}{}^2-\tau_1P_s^1+\tau_2P_s^3\ , \qquad &
\frac{\rmd P_s^3}{\rmd \tau_U} =F^{\rm (sc)}{}^3-\tau_2P_s^2\ ,
\end{eqnarray}
where the spin-curvature force $F^{\rm (sc)}$ is orthogonal to $U$.

In the case in which $U$ is a general circular orbit in a stationary axisymmetric spacetime the FS frame is known \cite{iyer-vish}, and the FS curvature and torsions are constant along $U$. 
Furthermore, in reflection-symmetric stationary axisymmetric spacetimes and for equatorial circular orbits the second torsion vanishes. The whole set of equations of motion (\ref{FSspineqsgen}) and (\ref{FSeqsmotogen}) thus reduces to
\begin{eqnarray}
\label{FSeqsgenfin}
\frac{\rmd S_{12}}{\rmd \tau_U}&=&\kappa S_{02}\ ,\quad
\frac{\rmd S_{13}}{\rmd \tau_U} = \tau_1 S_{23}+\kappa S_{03}\ ,\quad
\frac{\rmd S_{23}}{\rmd \tau_U} = -\tau_1 S_{13}\ , \nonumber\\
\quad
0&=&\frac{\rmd m}{\rmd \tau_U}+\kappa\frac{\rmd S_{01}}{\rmd \tau_U} -\kappa\tau_1S_{02}\ , \nonumber\\
\frac{\rmd^2 S_{01}}{\rmd \tau_U^2}&=&F^{\rm (sc)}{}^1-m\kappa+2\tau_1\frac{\rmd S_{02}}{\rmd \tau_U}+\tau_1 [\tau_1S_{01}-\kappa S_{12}]\ , \nonumber\\
\frac{\rmd^2 S_{02}}{\rmd \tau_U^2}&=&F^{\rm (sc)}{}^2-2\tau_1\frac{\rmd S_{01}}{\rmd \tau_U}+(\kappa^2+\tau_1^2)S_{02}\ , \nonumber\\
\frac{\rmd^2 S_{03}}{\rmd \tau_U^2}&=&F^{\rm (sc)}{}^3+\kappa[\tau_1 S_{23}+\kappa S_{03}]\ .
\end{eqnarray}
Once this system of equations is solved, the spatial momentum components $P_s$ may be expressed as 
\beq\fl\quad
\label{Pscompts}
P_s{}_1=\frac{\rmd S_{01}}{\rmd \tau_U}-\tau_1S_{02}\ , \quad
P_s{}_2=\frac{\rmd S_{02}}{\rmd \tau_U}+\tau_1S_{01}-\kappa S_{12}\ , \quad
P_s{}_3=\frac{\rmd S_{03}}{\rmd \tau_U}-\kappa S_{13}\ .
\eeq
To solve this system we need the explicit components of the spin-curvature force $F^{\rm (sc)}$, which couples the background curvature to the components of the spin tensor.

\section{Circular orbits in the Kerr spacetime}

The Kerr metric  in standard Boyer-Lindquist coordinates is given by
\begin{eqnarray}
\rmd s^2 &=& -\left(1-\frac{2Mr}{\Sigma}\right)\rmd t^2 -\frac{4aMr}{\Sigma}\sin^2\theta\rmd t\rmd\phi+ \frac{\Sigma}{\Delta}\rmd r^2 +\Sigma\rmd \theta^2\nonumber\\
&&+\frac{(r^2+a^2)^2-\Delta a^2\sin^2\theta}{\Sigma}\sin^2 \theta \rmd \phi^2\ ,
\end{eqnarray}
where $\Delta=r^2-2Mr+a^2$ and $\Sigma=r^2+a^2\cos^2\theta$; here $a$ and $M$ are the specific angular momentum and total mass of the spacetime solution. The event horizon and inner horizon are located at $r_\pm=M\pm\sqrt{M^2-a^2}$. 

Introduce the zero angular momentum observer (ZAMO) family of fiducial observers, with four-velocity
\beq
\label{n}
n=N^{-1}(\partial_t-N^{\phi}\partial_\phi)\ ,
\eeq
where $N=(-g^{tt})^{-1/2}$ and $N^{\phi}=g_{t\phi}/g_{\phi\phi}$ are the lapse and shift functions respectively. The natural orthonormal frame adapted to the ZAMOs is given by
\beq
\label{zamoframe}
e_{\hat t}=n , \,\quad
e_{\hat r}=\frac1{\sqrt{g_{rr}}}\partial_r, \,\quad
e_{\hat \theta}=\frac1{\sqrt{g_{\theta \theta }}}\partial_\theta, \,\quad
e_{\hat \phi}=\frac1{\sqrt{g_{\phi \phi }}}\partial_\phi .
\eeq

The 4-velocity $U$ of a particle uniformly rotating on circular orbits
can be parametrized either by the (constant) angular velocity with respect to infinity $\zeta$ or equivalently by the (constant) linear velocity  $\nu$ with respect to the ZAMOs 
\beq
\label{orbita}
U=\Gamma [\partial_t +\zeta \partial_\phi ]=\gamma [e_{\hat t} +\nu e_{\hat \phi}], \qquad \gamma=(1-\nu^2)^{-1/2}\ ,
\eeq
where
\beq\fl\qquad
\Gamma =\left[ N^2-g_{\phi\phi}(\zeta+N^{\phi})^2 \right]^{-1/2} = \gamma/N
\ ,\quad
\zeta=-N^{\phi}+(g_{\phi\phi})^{-1/2} N  \nu \ .
\eeq
We limit our analysis to  the equatorial plane ($\theta=\pi/2$) of the Kerr solution.
Note that both $\theta=\pi/2$ and $r=r_0$ are constants along any given circular orbit, and that the azimuthal coordinate along the orbit depends on the coordinate time $t$ or proper time $\tau$ along that orbit according to 
\beq\label{eq:phitau}
  \phi -\phi_0 = \zeta t = \Omega_U \tau_U \ ,\quad
\Omega_U =\Gamma\zeta
\ ,
\eeq
defining the corresponding coordinate and proper time orbital angular velocities $\zeta$ and $\Omega_U$. These determine the rotation of the spherical frame with respect to a nonrotating frame at infinity. 

On the equatorial plane of the Kerr solution there exist many special circular orbits  \cite{bjdf,idcf1, idcf2,bjm}. The co-rotating $(+)$ and counter-rotating $(-)$ timelike circular geodesics are particularly interesting, with angular and linear velocities respectively given by
\beq\fl\quad
\zeta_{({\rm geo})\, \pm}\equiv\zeta_{\pm}=\left[a\pm (M/r^3)^{1/2}\right]^{-1}\ , \quad 
\nu_{({\rm geo})\, \pm}\equiv \nu_\pm =\frac{a^2\mp2a\sqrt{Mr}+r^2}{\sqrt{\Delta}(a\pm r\sqrt{r/M})}\ .
\eeq 
The corresponding timelike conditions $|\nu_\pm|<1$ identify the allowed regions for the radial coordinate where co/counter-rotating geodesics exist: $r>r_{{(\rm geo)}\pm}$, where 
\beq
r_{{(\rm geo)}\pm}=2M\left\{1+\cos\left[\frac23\arccos\left(\pm\frac{a}{M}\right)\right]\right\}\ .
\eeq
The  ``geodesic meeting point observers'' defined in \cite{idcf2} have
\beq
\nu_{\rm (gmp)}=\frac{
\nu_{+}+\nu_{-}}{2}=-\frac{aM(3r^2+a^2)}{\sqrt{\Delta}(r^3-a^2M)}\ .
\eeq
It is convenient to introduce the Lie relative curvature of each orbit \cite{idcf2}
\beq
k_{\rm (lie)}=-\partial_{\hat r} \ln \sqrt{g_{\phi\phi}}=-\frac{(r^3-a^2M)\sqrt{\Delta}}{r^2(r^3+a^2r+2a^2M)}\ ,
\eeq
as well as
a natural Frenet-Serret frame along $U$ \cite{iyer-vish}
\begin{eqnarray}\fl\quad
E_0 \equiv U=\gamma [n+\nu e_{\hat \phi}]\ , \quad 
E_1 =e_{\hat r}\ , \quad 
E_2 \equiv E_{\hat \phi}=\gamma [\nu n+e_{\hat \phi}]\ , \quad  
E_3 = -e_{\hat\theta}\ . 
\end{eqnarray} 
The second torsion $\tau_2$ is zero for equatorial plane circular orbits in the Kerr spacetime; moreover, the geodesic curvature $\kappa$ and the first torsion $\tau_1$ are simply related by
\beq
\tau_1= -\frac{1}{2\gamma^2} \frac{\rmd \kappa}{\rmd \nu}\ ,
\eeq
so that 
\beq\fl
\label{ketau1}
\kappa = k_{\rm (lie)}\gamma^2 (\nu-\nu_+)(\nu-\nu_-)\ ,\quad
\tau_1 = k_{\rm (lie)}\gamma^2 \nu_{\rm (gmp)} (\nu-\nu_{{\rm (crit)}+})(\nu-\nu_{{\rm (crit)}-})\ ,
\eeq
where
\begin{eqnarray}
\fl\quad\nu_{{\rm (crit)}\pm}&=&\frac{\gamma_- \nu_- \mp \gamma_+ \nu_+}{\gamma_- \mp \gamma_+} \nonumber \\
\fl\quad&=& -\frac1{2Ma(3r^2+a^2)\sqrt{\Delta}}\Big[-2a^2M(a^2-3Mr)+r^2(r^2+a^2)(r-3M)\nonumber\\
\fl\quad &&\pm(r^3+a^2r+2a^2M)\sqrt{r}\sqrt{r(r-3M)^2-4a^2M}\Big]\
\end{eqnarray}
and $\nu_{{\rm (ext)}}\equiv\nu_{{\rm (crit)}-}$
is the velocity of the ``extremely accelerated observers'' \cite{idcf2,fdfacc} at which the first torsion vanishes. This is a critical value for the curvature $\kappa$ as a function of $\nu$ which corresponds to an extreme value of the curvature.
The corresponding 4-velocity is timelike in the regions $r_+<r<r_{{(\rm geo)}+}$ and $r>r_{{(\rm geo)}-}$.

\section{Solving the equations of motion: preliminary steps}

Now specialize the entire set of evolution equations (\ref{FSeqsgenfin}) to the case of a spinning test particle moving along circular orbits in the equatorial plane of a Kerr black hole.
The only nonvanishing Frenet-Serret frame components of the Riemann tensor are 
\begin{eqnarray}
\label{riemFS}
R_{0101}&=-R_{2323}=E_{11}\ , \quad & R_{0202}=-R_{1313}=E_{22}\ , \nonumber\\
R_{0303}&=-R_{1212}=E_{33}\ , \quad & R_{0112}=R_{0323}=H_{13}\ ,
\end{eqnarray}
where the electric and magnetic parts of the Weyl tensor have been introduced.
They are related to the  ZAMO frame components (hatted indices) by
\begin{eqnarray}\fl\quad
E_{11}&=\gamma^2[E_{\hat r \hat r}-\nu(2H_{\hat r \hat \theta}+\nu E_{\hat \theta \hat \theta})]\ , \quad & E_{22}=E_{\hat \phi \hat \phi}\ , \nonumber\\
\fl\quad
E_{33}&=\gamma^2[E_{\hat \theta \hat \theta}+\nu(2H_{\hat r \hat \theta}-\nu E_{\hat r \hat r})]\ , \quad & H_{13}=\gamma^2[(1+\nu^2)H_{\hat r \hat \theta}-\nu(E_{\hat r \hat r}-E_{\hat \theta \hat \theta})]\ ,
\end{eqnarray}
where
\begin{eqnarray}\fl\quad
E_{\hat r \hat r} &=& -E_{\hat \theta \hat \theta}-\frac{M}{r^3} 
= R_{\hat t\hat r\hat t\hat r} = -R_{\hat\theta \hat\phi \hat\theta \hat\phi}\ , 
\quad
E_{\hat \phi \hat \phi} = \frac{M}{r^3}
= R_{\hat t \hat\phi \hat t \hat\phi} = -R_{\hat r \hat\theta \hat r \hat\theta}\ , 
\nonumber\\\fl\quad
E_{\hat \theta \hat \theta}&=&\frac{M}{r^4}\frac{(r^2+a^2)(r^2+3a^2)-4a^2Mr}{r^3+a^2r+2a^2M}
= R_{\hat t \hat\theta \hat t \hat\theta} = -R_{\hat r \hat\phi \hat r \hat\phi}\ , \nonumber\\\fl\quad 
H_{\hat r \hat \theta} &=& -\frac{3Ma\Delta^{1/2}}{r^4}\frac{r^2+a^2}{r^3+a^2r+2a^2M}
= R_{\hat t\hat r\hat r\hat \phi}=- R_{\hat t \hat\theta \hat\theta \hat\phi}\ .
\end{eqnarray}

Eq.~(\ref{riemFS}) implies that the components of the spin-curvature-coupling force are given by
\beq\fl\quad
F^{\rm (sc)}{}^1=H_{13}S_{12}-E_{11}S_{01}\ , \,\, 
F^{\rm (sc)}{}^2=-E_{22}S_{02}\ , \,\,
F^{\rm (sc)}{}^3=H_{13}S_{23}-E_{33}S_{03}\ . 
\eeq
Hence Eqs.~(\ref{FSeqsgenfin}) become
\begin{eqnarray}
\label{KerrFSeq1}
\frac{\rmd S_{12}}{\rmd \tau_U}&=&\kappa S_{02}\ , \\
\label{KerrFSeq2}
\frac{\rmd S_{13}}{\rmd \tau_U}&=&\tau_1 S_{23}+\kappa S_{03}\ , \\
\label{KerrFSeq3}
\frac{\rmd S_{23}}{\rmd \tau_U}&=&-\tau_1 S_{13}\ , \\
\label{KerrFSeq4}
\quad
0&=&\frac{\rmd m}{\rmd \tau_U}+\kappa\frac{\rmd S_{01}}{\rmd \tau_U} -\kappa\tau_1S_{02}\ , \\
\label{KerrFSeq5}
\frac{\rmd^2 S_{01}}{\rmd \tau_U^2}&=&2\tau_1\frac{\rmd S_{02}}{\rmd \tau_U}-m\kappa+(\tau_1^2-E_{11})S_{01} -(\kappa\tau_1-H_{13})S_{12}\ , \\
\label{KerrFSeq6}
\frac{\rmd^2 S_{02}}{\rmd \tau_U^2}&=&-2\tau_1\frac{\rmd S_{01}}{\rmd \tau_U}+(\kappa^2+\tau_1^2-E_{22})S_{02}\ , \\
\label{KerrFSeq7}
\frac{\rmd^2 S_{03}}{\rmd \tau_U^2}&=&(\kappa^2-E_{33})S_{03}+(\kappa\tau_1+H_{13})S_{23}\ .
\end{eqnarray}
Once this system of constant coefficient linear differential equations is solved for $m$ and the spin tensor components, one may then evaluate $P$.
This system may be decoupled, leading to solutions which are either exponentials or sinusoidals or polynomial functions of the proper time. The elimination method for decoupling the equations is crucially different depending on whether $\kappa=0$ (geodesic motion) or $\tau_1=0$ (extremely accelerated motion, corresponding to extreme values of the curvature $\kappa$ as a function of the velocity) or neither condition holds, and so must be considered separately.

The projection of the spin tensor into the local rest space of $U$ (i.e., the subspace of the tangent space orthogonal to $U$) defines the spin vector by spatial duality
\beq
S^\beta={\textstyle\frac12} \eta_\alpha{}^{\beta\gamma\delta}U^\alpha S_{\gamma\delta}\ ,
\eeq
where $\eta_{\alpha\beta\gamma\delta}=\sqrt{-g} \epsilon_{\alpha\beta\gamma\delta}$ is the volume 4-form and $\epsilon_{\alpha\beta\gamma\delta}$ ($\epsilon_{\hat t\hat r\hat\theta\hat\phi}=1$) is the Levi-Civita alternating symbol. Its Frenet-Serret frame components are 
\beq
\label{spinFS}
(S^1,S^2,S^3)=(S_{23},-S_{13},S_{12}) \ .
\eeq
It is useful to introduce the spin scalar
\beq
\label{sinv}
s^2
=\frac12 S_{\mu\nu}S^{\mu\nu}
=-S_{01}^2 -S_{02}^2 -S_{03}^2+S_{12}^2 +S_{13}^2+S_{23}^2\ . 
\eeq
In general $s$ is not constant along the trajectory of a spinning particle. 
The requirement which is essential to the validity of the Mathisson-Papapetrou model and the test particle approach is that the characteristic length scale $|s|/m$ associated with the particle's internal structure be small compared to the natural length scale $M$ associated with the background field \cite{mol}. Hence the following condition must be assumed to hold: $|s|/(mM)\ll 1$.

The Pirani conditions in the FS frame are simply 
\beq
\label{PcondFS}
S_{0a}=0\ .
\eeq
The Tulczyjew conditions are instead
\beq
\label{TcondFS}
-m S_{0b}\delta^b_\alpha+S_{\alpha b}P_s{}_b=0\ ,
\eeq
and replacing $P_s$ by its equivalent spin expression, they become
\begin{eqnarray}\fl\quad
\label{Tconds}
0&=&S_{01}\frac{\rmd S_{01}}{\rmd \tau_U}+S_{02}\frac{\rmd S_{02}}{\rmd \tau_U}+S_{03}\frac{\rmd S_{03}}{\rmd \tau_U}-\kappa(S_{02}S_{12}+S_{03}S_{13})\ , \nonumber\\
\fl\quad
0&=&S_{12}\frac{\rmd S_{02}}{\rmd \tau_U}+S_{13}\frac{\rmd S_{03}}{\rmd \tau_U}-\kappa(S_{12}^2+S_{13}^2)-(m-\tau_1S_{12})S_{01}\ , \nonumber\\
\fl\quad
0&=&-S_{12}\frac{\rmd S_{01}}{\rmd \tau_U}+S_{23}\frac{\rmd S_{03}}{\rmd \tau_U}-(m-\tau_1S_{12})S_{02}-\kappa S_{13}S_{23}\ , \nonumber\\
\fl\quad
0&=&S_{13}\frac{\rmd S_{01}}{\rmd \tau_U}+S_{23}\frac{\rmd S_{02}}{\rmd \tau_U}+mS_{03}-\kappa S_{12}S_{23}+\tau_1(S_{01}S_{23}-S_{02}S_{13})\ .
\end{eqnarray}

We are now ready to discuss the solutions of the equations of motion for the components of the spin tensor as well as the mass $m$ of the spinning particle, starting from the special cases of geodesic and extremely accelerated motion.
Although the FS procedure to construct the frame adapted to $U$ fails in these cases, the corresponding FS frames can be obtained by taking the respective limits $\kappa\to0$ and $\tau_1\to0$ of the general case.  

\section{Geodesic motion: the $\kappa=0$ case}

Eqs.~(\ref{KerrFSeq1})--(\ref{KerrFSeq7}) reduce to
\begin{eqnarray}
\label{KerrFSeq1geo}
S_{12}&=&c_0\ , \\
\label{KerrFSeq2geo}
\frac{\rmd S_{13}}{\rmd \tau_U}&=&\tau_1 S_{23}\ , \\
\label{KerrFSeq3geo}
\frac{\rmd S_{23}}{\rmd \tau_U}&=&-\tau_1 S_{13}\ , \\
\label{KerrFSeq4geo}
m&=&c_m\ , \\
\label{KerrFSeq5geo}
\frac{\rmd^2 S_{01}}{\rmd \tau_U^2}&=&2\tau_1\frac{\rmd S_{02}}{\rmd \tau_U}+(\tau_1^2-E_{11})S_{01} +H_{13}c_0\ , \\
\label{KerrFSeq6geo}
\frac{\rmd S_{02}}{\rmd \tau_U}&=&c_1-2\tau_1S_{01}\ , \\
\label{KerrFSeq7geo}
\frac{\rmd^2 S_{03}}{\rmd \tau_U^2}&=&-E_{33}S_{03}+H_{13}S_{23}\ ,
\end{eqnarray}
where the first torsion as well as the electric and magnetic parts of the Weyl tensor are evaluated at $\nu=\nu_\pm$, and the important relation $\tau_1^2=E_{22}$ found by explicit calculation has been used.
The cases $\nu=\nu_+$ and $\nu=\nu_-$ must be considered separately.

Using Eq. (\ref{KerrFSeq6geo}) in Eq. (\ref{KerrFSeq5geo}) we obtain
\beq
\label{KerrFSeq5geob}
0=\frac{\rmd^2 S_{01}}{\rmd \tau_U^2}+\omega_{g\pm}^2S_{01} -H_{13}c_0-2\tau_1c_1\ ,
\eeq
where 
\beq
\omega_{g\pm}^2=3\tau_1^2+E_{11}=\frac{M}{r^3}\left[1-\frac{3(a\mp\sqrt{Mr})^2}{r^2-3Mr\pm2a\sqrt{Mr}}\right]\ .
\eeq
Equation (\ref{KerrFSeq5geob}) is easily solved, and the corresponding solution for $S_{02}$ comes from Eq. (\ref{KerrFSeq6geo}). The character of the solution depends on whether $\omega_{g\pm}$ is real or imaginary or zero. 
It is easy to show that for a fixed value of the black hole rotation parameter there exists only one real root $r={\bar r}_\pm$ of the equation $\omega_{g\pm}=0$, with $\omega_{g\pm}^2<0$ $(>0)$ for $r<{\bar r}_\pm$ $(>{\bar r}_\pm)$.
Hence we have the following three cases. 

\begin{enumerate}

\item $\omega_{g\pm}^2>0$:

\begin{eqnarray}\fl\quad
\label{solkappaeq0I}
S_{01}&=& c_2\cos\omega_{g\pm}\tau+c_3\sin\omega_{g\pm}\tau+\frac{H_{13}c_0+2\tau_1c_1}{\omega_{g\pm}^2}\ , \nonumber\\
\fl\quad
S_{02}&=&c_4-\frac{2\tau_1H_{13}c_0+(\tau_1^2-E_{11})c_1}{\omega_{g\pm}^2}\tau-\frac{2\tau_1}{\omega_{g\pm}}[c_2\sin\omega_{g\pm}\tau-c_3\cos\omega_{g\pm}\tau]\ ; 
\end{eqnarray}

\item $\omega_{g\pm}^2<0$ (imaginary frequency):

\begin{eqnarray}\fl\quad
\label{solkappaeq0II}
S_{01}&=& c_2 e^{-{\bar \omega}_{g\pm}\tau}+c_3 e^{{\bar \omega}_{g\pm}\tau}-\frac{H_{13}c_0+2\tau_1c_1}{{\bar \omega}_{g\pm}^2}\ , \nonumber\\
\fl\quad
S_{02}&=&c_4+\frac{2\tau_1H_{13}c_0+(\tau_1^2-E_{11})c_1}{{\bar \omega}_{g\pm}^2}\tau+\frac{2\tau_1}{{\bar \omega}_{g\pm}}[c_2e^{-{\bar \omega}_{g\pm}\tau}-c_3e^{{\bar \omega}_{g\pm}\tau}]\ , 
\end{eqnarray}
where $\omega_{g\pm}=i{\bar \omega}_{g\pm}$;

\item $\omega_{g\pm}=0$:

\begin{eqnarray}\fl\quad
\label{solkappaeq0III}
S_{01}&=&c_2+c_3\tau+(H_{13}c_0+2\tau_1c_1)\frac{\tau^2}{2}\ , \nonumber\\
\fl\quad
S_{02}&=&c_4+(c_1-2\tau_1c_2)\tau-\tau_1c_3\tau^2-\tau_1(H_{13}c_0+2\tau_1c_1)\frac{\tau^3}{3}\ . 
\end{eqnarray}

\end{enumerate}

To solve for the remaining components of the spin tensor,
take the $\tau_U$ derivative of Eq.~(\ref{KerrFSeq3geo}) and use Eqs.~(\ref{KerrFSeq2geo}) to get
\beq
\label{KerrFSeq3geob}
\frac{\rmd^2 S_{23}}{\rmd \tau_U^2}+\tau_1^2S_{23}=0\ ,
\eeq
which can be easily integrated; Eqs.~(\ref{KerrFSeq3geo}) and (\ref{KerrFSeq7geo}) give the corresponding solutions for $S_{13}$ and $S_{03}$. We then obtain
\begin{eqnarray}\fl\quad
\label{solkappaeq0altre3}
S_{23}&=& c_5\cos\tau_1\tau+c_6\sin\tau_1\tau\ ,\quad
S_{13} = c_5\sin\tau_1\tau-c_6\cos\tau_1\tau\ , \nonumber\\
\fl\quad
S_{03}&=&c_7\cos\sqrt{E_{33}^\pm}\tau+c_8\sin\sqrt{E_{33}^\pm}\tau-\frac{H_{13}}{\tau_1^2-E_{33}^\pm}[c_5\cos\tau_1\tau+c_6\sin\tau_1\tau]\ , 
\end{eqnarray}
where both quantities 
\beq
\tau_1^2=\frac{M}{r^3}\ , \quad 
E_{33}^\pm=\frac{M}{r^3}\left[1+\frac{3(a\mp\sqrt{Mr})^2}{r^2-3Mr\pm2a\sqrt{Mr}}\right]\ ,
\eeq
are always positive in the allowed regions in both cases $\nu=\nu_+$ and $\nu=\nu_-$.

At this point the supplementary conditions impose constraints on the constants of integration which appear in the solution.
The Pirani conditions (\ref{PcondFS}) lead only to the trivial solution
\beq
c_0=c_1=c_2=c_3=c_4=c_5=c_6=c_7=c_8=0\ , 
\eeq
where $c_m$ is the \lq\lq constant" mass of the particle (\ref{KerrFSeq4geo}). In other words all the components of the spin tensor must be zero, which means that a non-zero spin is incompatible with geodesic motion for a spinning particle in circular motion. 

The Tulczyjew supplementary conditions (\ref{TcondFS}) imply 
\begin{eqnarray}
\label{Tcondsgeo}
0&=&S_{01}\frac{\rmd S_{01}}{\rmd \tau_U}+S_{03}\frac{\rmd S_{03}}{\rmd \tau_U}+S_{02}(c_1-2\tau_1S_{01})\ , \nonumber\\
0&=&S_{13}\frac{\rmd S_{03}}{\rmd \tau_U}-(c_m+\tau_1c_0)S_{01}+c_0c_1\ , \nonumber\\
0&=&c_0\frac{\rmd S_{01}}{\rmd \tau_U}-S_{23}\frac{\rmd S_{03}}{\rmd \tau_U}+(c_m-\tau_1c_0)S_{02}\ , \nonumber\\
0&=&S_{13}\left[\frac{\rmd S_{01}}{\rmd \tau_U}-\tau_1S_{02}\right]+S_{23}(c_1-\tau_1S_{01})+c_m S_{03}\ .
\end{eqnarray}
Consider first the case $\omega_{g\pm}\not=0$.
By substituting the solutions given by Eqs.~(\ref{solkappaeq0I}), (\ref{solkappaeq0II}) and (\ref{solkappaeq0altre3}) into Eqs. (\ref{Tcondsgeo}), we obtain the following conditions. 
Either
\beq
c_0=c_1=c_2=c_3=c_4=c_5=c_6=c_7=c_8=0\ ,
\eeq
with $c_m=m$, corresponding to the zero spin case where geodesic motion is of course allowed, 
or
\beq\fl\quad
c_2=c_3=c_5=c_6=c_7=c_8=0\ , \qquad c_1=-\frac{2\tau_1H_{13}}{\tau_1^2-E_{11}}c_0\ , \qquad c_m=\tau_1c_0\ ,
\eeq
with $c_0$, $c_4$ arbitrary, implying that the only nonvanishing components of the spin tensor are
\beq
S_{01}=-\frac{H_{13}}{\tau_1^2-E_{11}}\frac{c_m}{\tau_1}\ , \qquad S_{02}=c_4\ , \qquad S_{12}=S^3=\frac{c_m}{\tau_1}\ . 
\eeq
Finally consider the remaining case $\omega_{g\pm}=0$.
By substituting into Eqs. (\ref{Tcondsgeo}) the solutions given by Eqs.~(\ref{solkappaeq0III}) and (\ref{solkappaeq0altre3}), we obtain the following conditions: 
either
\beq
c_0=c_1=c_2=c_3=c_4=c_5=c_6=c_7=c_8=0\ ,
\eeq
with $c_m=m$, as expected,  
or
\beq\fl\quad
c_3=c_5=c_6=c_7=c_8=0\ , \quad c_1=-\frac{H_{13}}{2\tau_1}c_0\ , \quad c_2=\frac{c_1}{2\tau_1}\ , \quad c_m=\tau_1c_0\ ,
\eeq
with $c_0$, $c_4$ arbitrary, implying that the only nonvanishing components of the spin tensor are
\beq
S_{01}=-\frac{H_{13}}{4\tau_1^3}c_m\ , \qquad S_{02}=c_4\ , \qquad S_{12}=S^3=\frac{c_m}{\tau_1}\ . 
\eeq

Thus if the center of mass of the test particle is constrained to be a circular geodesic, either the spin tensor is identically zero or the spin vector has a nonzero component $S^3$ out of the plane of the orbit, having a constant value fixed by the particle mass. 

\section{Extremely accelerated motion: the $\tau_1=0$ case}

Imposing the condition $\tau_1=0$, Eqs.~(\ref{KerrFSeq1})--(\ref{KerrFSeq7}) reduce to
\begin{eqnarray}
\label{KerrFSeq1crit}
\frac{\rmd S_{12}}{\rmd \tau_U}&=&\kappa S_{02}\ , \\
\label{KerrFSeq2crit}
\frac{\rmd S_{13}}{\rmd \tau_U}&=&\kappa S_{03}\ , \\
\label{KerrFSeq3crit}
S_{23}&=&c_0\ , \\
\label{KerrFSeq4crit}
m&=&c_1-\kappa S_{01}\ , \\
\label{KerrFSeq5crit}
\frac{\rmd^2 S_{01}}{\rmd \tau_U^2}&=&-m\kappa-E_{11}S_{01} +H_{13}S_{12}\ , \\
\label{KerrFSeq6crit}
\frac{\rmd^2 S_{02}}{\rmd \tau_U^2}&=&(\kappa^2-E_{22})S_{02}\ , \\
\label{KerrFSeq7crit}
\frac{\rmd^2 S_{03}}{\rmd \tau_U^2}&=&(\kappa^2-E_{33})S_{03}+H_{13}c_0\ ,
\end{eqnarray}
where the FS curvature as well as the electric and magnetic parts of the Weyl tensor are evaluated at $\nu=\nu_{{\rm (ext)}}$.

Eqs.~(\ref{KerrFSeq6crit}) and (\ref{KerrFSeq7crit}) are easily solved; the corresponding solutions for $S_{12}$ and $S_{13}$ follow immediately from Eqs. (\ref{KerrFSeq1crit}) and (\ref{KerrFSeq2crit}).
Using Eq. (\ref{KerrFSeq4crit}) in Eq. (\ref{KerrFSeq5crit}) leads to
\beq
\label{KerrFSeq5critb}
\frac{\rmd^2 S_{01}}{\rmd \tau_U^2}=(\kappa^2-E_{11})S_{01} +H_{13}S_{12}-\kappa c_1\ ,
\eeq
which can be easily integrated as well.
The character of the solutions critically depends on the sign of the following quantities in the allowed regions $r_+<r<r_{{(\rm geo)}+}$ and $r>r_{{(\rm geo)}-}$:
\begin{eqnarray}\fl\quad
\omega_{\rm ext{}1}^2=-\kappa^2+E_{11}&=&
\frac{1}{2r^3\Delta}\bigg[\frac{3M}{\xi(r)}(Mr-a^2)(3r^2-5Mr+2a^2)\nonumber\\
\fl\quad
&&-3M\Delta\left(1+\frac{4Mr}{\xi(r)}\right)+(r-4M)\xi(r)-\frac{\xi(r)^2}{r}\bigg]\ , \nonumber\\
\fl\quad
\omega_{\rm ext{}2}^2=-\kappa^2+E_{22}&=&
\frac{\xi(r)}{2r^3\Delta}\left[r-M-\frac{\xi(r)}{r}\right]\ , \nonumber\\
\fl\quad
\omega_{\rm ext{}3}^2=-\kappa^2+E_{33}&=&
\omega_{\rm ext{}1}^2+\frac{3M}{r^3\xi(r)}(r^2-3Mr+a^2)\ ,
\end{eqnarray}
where $\xi(r)=r^{1/2}[r^3-6Mr^2+9M^2r-4Ma^2]^{1/2}$.
It is easy to show that for every fixed value of the black hole rotation parameter the quantity $\omega_{\rm ext{}2}^2$ is always positive in the allowed region, so that the corresponding solution for $S_{02}$ will be oscillating everywhere:
\beq
\label{sols02tau1eq0}
S_{02}= c_2\cos\omega_{\rm ext{}2}\tau+c_3\sin\omega_{\rm ext{}2}\tau\ .
\eeq
Eq.~(\ref{KerrFSeq1crit}) then gives 
\beq
\label{sols12tau1eq0}
S_{12}= c_4-\frac{\kappa}{\omega_{\rm ext{}2}}[c_3\cos\omega_{\rm ext{}2}\tau-c_2\sin\omega_{\rm ext{}2}\tau]\ .
\eeq
The quantity $\omega_{\rm ext{}1}^2$ is positive in the region $r_+<r<r_{{(\rm geo)}+}$ and negative for $r>r_{{(\rm geo)}-}$, and viceversa for $\omega_{\rm ext{}3}^2$. 
Thus we have

\begin{enumerate}

\item $r_+<r<r_{{(\rm geo)}+}$:

\begin{eqnarray}\fl\quad
\label{soltau1eq0I}
S_{01}&=& c_5\cos\omega_{\rm ext{}1}\tau+c_6\sin\omega_{\rm ext{}1}\tau+\frac{H_{13}c_4-\kappa c_1}{\omega_{\rm ext{}1}^2}\nonumber\\
\fl\quad
&&-\frac{\kappa H_{13}}{\omega_{\rm ext{}2}(E_{11}-E_{22})}[c_3\cos\omega_{\rm ext{}2}\tau-c_2\sin\omega_{\rm ext{}2}\tau]\ , \nonumber\\
\fl\quad
S_{03}&=&c_7 e^{-{\bar \omega}_{\rm ext{}3}\tau}+c_8 e^{{\bar \omega}_{\rm ext{}3}\tau}-\frac{H_{13}c_0}{{\bar \omega}_{\rm ext{}3}^2}\ , \nonumber\\
\fl\quad
S_{13}&=&c_9-\frac{\kappa H_{13}c_0}{{\bar \omega}_{\rm ext{}3}^2}\tau-\frac{\kappa}{{\bar \omega}_{\rm ext{}3}}[c_7 e^{-{\bar \omega}_{\rm ext{}3}\tau}-c_8 e^{{\bar \omega}_{\rm ext{}3}\tau}]\ ,  
\end{eqnarray}
where $\omega_{\rm ext{}3}=i{\bar \omega}_{\rm ext{}3}$.

\item $r>r_{{(\rm geo)}-}$:

\begin{eqnarray}\fl\quad
\label{soltau1eq0II}
S_{01}&=& c_5 e^{-{\bar \omega}_{\rm ext{}1}\tau}+c_6 e^{{\bar \omega}_{\rm ext{}1}\tau}-\frac{H_{13}c_4-\kappa c_1}{{\bar \omega}_{\rm ext{}1}^2}\nonumber\\
\fl\quad
&&+\frac{\kappa H_{13}}{\omega_{\rm ext{}2}({\bar \omega}_{\rm ext{}1}^2+\omega_{\rm ext{}2}^2)}[c_3\cos\omega_{\rm ext{}2}\tau-c_2\sin\omega_{\rm ext{}2}\tau]\ , \nonumber\\
\fl\quad
S_{03}&=&c_7 \cos\omega_{\rm ext{}3}\tau+c_8 \sin\omega_{\rm ext{}3}\tau+\frac{H_{13}c_0}{\omega_{\rm ext{}3}^2}\ , \nonumber\\
\fl\quad
S_{13}&=&c_9+\frac{\kappa H_{13}c_0}{\omega_{\rm ext{}3}^2}\tau-\frac{\kappa}{\omega_{\rm ext{}3}}[c_8 \cos\omega_{\rm ext{}3}\tau-c_7 \sin\omega_{\rm ext{}3}\tau]\ .  
\end{eqnarray}

\end{enumerate}

Next impose the standard supplementary conditions.
The Pirani conditions (\ref{PcondFS}) imply
\beq
c_0=c_2=c_3=c_5=c_6=c_7=c_8=0\ , \qquad c_4=\frac{\kappa}{H_{13}}c_1\ ,
\eeq
with $c_1=m$ and $c_9$ arbitrary.
Thus the only nonvanishing components of the spin tensor are given by 
\beq
\label{S1213Psoltau1eq0}
S_{12}=S^3=\frac{m\kappa}{H_{13}}\ , \qquad S_{13}=-S^2=c_9\ .
\eeq

The Tulczyjew supplementary conditions (\ref{TcondFS}) imply instead
\begin{eqnarray}
\label{Tcondstau1eq0}
0&=&S_{01}\frac{\rmd S_{01}}{\rmd \tau_U}+S_{02}\frac{\rmd S_{02}}{\rmd \tau_U}+S_{03}\frac{\rmd S_{03}}{\rmd \tau_U}-\kappa(S_{02}S_{12}+S_{03}S_{13})\ , \nonumber\\
0&=&S_{12}\frac{\rmd S_{02}}{\rmd \tau_U}+S_{13}\frac{\rmd S_{03}}{\rmd \tau_U}+\kappa(S_{01}^2-S_{12}^2-S_{13}^2)-c_1S_{01}\ , \nonumber\\
0&=&-S_{12}\frac{\rmd S_{01}}{\rmd \tau_U}+c_0\frac{\rmd S_{03}}{\rmd \tau_U}+\kappa (S_{01}S_{02}-c_0S_{13})-c_1S_{02}\ , \nonumber\\
0&=&S_{13}\frac{\rmd S_{01}}{\rmd \tau_U}+c_0\frac{\rmd S_{02}}{\rmd \tau_U}-\kappa(S_{01}S_{03}+c_0S_{12})+c_1S_{03}\ .
\end{eqnarray}
Substituting into Eqs. (\ref{Tcondstau1eq0}) the solutions given by Eqs.~(\ref{sols02tau1eq0}), (\ref{sols12tau1eq0}) and (\ref{soltau1eq0I}) (holding in the region $r_+<r<r_{{(\rm geo)}+}$) leads to 
\beq
c_0=c_2=c_3=c_5=c_6=c_7=c_8=0\ ,
\eeq
with the remaining integration constants $c_1$, $c_4$, $c_9$ having to satisfy the condition  
\beq
0=\frac{m}{E_{11}}(\kappa m-H_{13}c_4)-\kappa(c_4^2+c_9^2)\ , 
\eeq
where
\beq
c_1=\frac{m\omega_{\rm ext{}1}^2+\kappa H_{13}c_4}{E_{11}}\ , 
\eeq
implying that the only nonvanishing components of the spin tensor are
\beq\fl\quad
S_{01}=-\frac{\kappa m-H_{13}c_4}{E_{11}}\ , \qquad S_{12}=S^3=c_4\ , \qquad S_{13}=-S^2=c_9\ . 
\eeq
The same result is obtained if the solution (\ref{soltau1eq0II}) valid in the region $r>r_{{(\rm geo)}-}$ is used instead of (\ref{soltau1eq0I}).
The value of the arbitrary constant $c_4$ can be fixed in such a way that $S_{01}=0$ ($c_4=\kappa m/H_{13}$), leading to the same solution as in the Pirani case (\ref{S1213Psoltau1eq0}).

Thus if the center of mass of the test particle follows an extremely accelerated circular orbit, the spin vector is allowed to have arbitrary constant values of the nonzero components $S^2$ and $S^3$, the latter depending on the particle mass. 

\section{The general case: $\nu\not= \nu_\pm$ and $\nu\not=\nu_{{\rm (ext)}}$}

Eqs.~(\ref{KerrFSeq1}) and (\ref{KerrFSeq4}) imply that 
\beq
\label{KerrFSeq4b}
m+\kappa S_{01}-\tau_1 S_{12}=c_m\ ,
\eeq
where $c_m$ is an arbitrary integration constant.
Using Eq. (\ref{KerrFSeq1}) in Eq. (\ref{KerrFSeq6}) leads to 
\beq
\label{KerrFSeq6b}
\frac{\rmd^2 S_{12}}{\rmd \tau_U^2}+2\kappa\tau_1S_{01}-(\kappa^2+\tau_1^2-E_{22})S_{12}=c_0\ .
\eeq
Using Eqs. (\ref{KerrFSeq1}) and (\ref{KerrFSeq4b}) in Eq. (\ref{KerrFSeq5}) leads to 
\beq\fl\quad
\label{KerrFSeq5b}
\frac{\rmd^2 S_{01}}{\rmd \tau_U^2}=2\frac{\tau_1}{\kappa}\frac{\rmd^2 S_{12}}{\rmd \tau_U^2}-\kappa c_m+ (\kappa^2+\tau_1^2-E_{11})S_{01} -(2\kappa\tau_1-H_{13})S_{12}\ .
\eeq
Then solving these last two equations for $\rmd^2S_{01}/\rmd \tau_U^2$ and $\rmd^2S_{12}/\rmd \tau_U^2$ gives
\begin{eqnarray}\fl\quad
\label{KerrFSeq5cKerrFSeq6c}
\frac{\rmd^2S_{12}}{\rmd \tau_U^2}&=&(\kappa^2+\tau_1^2-E_{22})S_{12}-2\kappa\tau_1S_{01} +c_0\ , \nonumber\\
\fl\quad
\frac{\rmd^2S_{01}}{\rmd \tau_U^2}&=&\left[2\frac{\tau_1}{\kappa}(\tau_1^2-E_{22})+H_{13}\right]S_{12}+ (\kappa^2-3\tau_1^2-E_{11}) S_{01}+2\frac{\tau_1}{\kappa}c_0-\kappa c_m\ .
\end{eqnarray}
An analogous pair of coupled equations is obtained for the components $S_{03}$ and $S_{23}$, taking the $\tau_U$ derivative of Eq.~(\ref{KerrFSeq3}) and using Eqs.~(\ref{KerrFSeq2}) and (\ref{KerrFSeq7}), namely
\begin{eqnarray}
\label{KerrFSeq3bKerrFSeq7b}
\frac{\rmd^2S_{23}}{\rmd \tau_U^2}&=&-\tau_1^2 S_{23}-\kappa\tau_1 S_{03}\ , \nonumber\\
\frac{\rmd^2S_{03}}{\rmd \tau_U^2}&=&(\kappa\tau_1+H_{13})S_{23}+(\kappa^2-E_{33})S_{03}\ .
\end{eqnarray}
The two pair of equations (\ref{KerrFSeq5cKerrFSeq6c}) and (\ref{KerrFSeq3bKerrFSeq7b}) can be easily solved. Then substituting these solutions into Eqs.~(\ref{KerrFSeq1}) and (\ref{KerrFSeq3}), one obtains the corresponding solutions for the remaining components of the spin tensor.

The solutions of the equations of motion for the components of the spin tensor and the mass $m$ of the spinning particle are
\begin{eqnarray}\fl\quad
\label{s23sol}
S_{23}&=&f_{\omega_-}+f_{\omega_+}\ , \\
\fl\quad
\label{s03sol}
S_{03}&=&-\frac{\omega_-^2+\tau_1^2}{\kappa\tau_1}f_{\omega_-}-\frac{\omega_+^2+\tau_1^2}{\kappa\tau_1}f_{\omega_+}\ , \\
\fl\quad
\label{s13sol}
S_{13}&=&-\frac{1}{\tau_1}\left[\frac{\rmd f_{\omega_-}}{\rmd\tau}+\frac{\rmd f_{\omega_+}}{\rmd\tau}\right]\ , \\
\fl\quad
\label{s12sol}
S_{12}&=&\frac{2\kappa^2\tau_1c_m-(\kappa^2+\tau_1^2-E_{11})c_0}{(\kappa^2-\tau_1^2)^2-(\kappa^2+\tau_1^2)(E_{11}+E_{22})+2\kappa\tau_1H_{13}+E_{11}E_{22}}\nonumber\\
\fl\quad
&&+f_{\Omega_-}+f_{\Omega_+}\ , \\
\fl\quad
\label{s01sol}
S_{01}&=&\frac{(2\kappa\tau_1-H_{13})c_0-\kappa c_m(\kappa^2+\tau_1^2-E_{22})}{(\kappa^2-\tau_1^2)^2-(\kappa^2+\tau_1^2)(E_{11}+E_{22})+2\kappa\tau_1H_{13}+E_{11}E_{22}}\nonumber\\
\fl\quad
&&-\frac{\Omega_-^2-(\kappa^2+\tau_1^2-E_{22})}{2\kappa\tau_1}f_{\Omega_-}-\frac{\Omega_+^2-(\kappa^2+\tau_1^2-E_{22})}{2\kappa\tau_1}f_{\Omega_+}\ , \\
\fl\quad
\label{s02sol}
S_{02}&=&\frac{1}{\kappa}\left[\frac{\rmd f_{\Omega_-}}{\rmd\tau}+\frac{\rmd f_{\Omega_+}}{\rmd\tau}\right]\ , \\ 
\fl\quad
\label{masssol}
m&=&\tau_1 S_{12}-\kappa S_{01}+c_m\ ,
\end{eqnarray}
where 
\beq
f_{\omega_\pm}=c_\pm e^{-\omega_\pm\tau}+d_\pm e^{\omega_\pm\tau}\ , \qquad
f_{\Omega_\pm}=h_\pm e^{-\Omega_\pm\tau}+l_\pm e^{\Omega_\pm\tau}\ ,
\eeq
$c_m, c_0, c_\pm, d_\pm, h_\pm, l_\pm$ are integration constants and
\begin{eqnarray}\fl\quad
\label{omegaspm}
\omega_{\pm}&=&\frac{\sqrt{2}}{2}\Big\{\kappa^2-\tau_1^2-E_{33}\nonumber\\
\fl\quad
&&\pm\left\{(\kappa^2-\tau_1^2)^2-[2(\kappa^2+\tau_1^2)-E_{33}]E_{33}-4\kappa\tau_1H_{13}\right\}^{1/2}\Big\}^{1/2}\ , \nonumber\\
\fl\quad
\Omega_{\pm}&=&\frac{\sqrt{2}}{2}\Big\{2(\kappa^2-\tau_1^2)-E_{22}-E_{11}\nonumber\\
\fl\quad
&&\pm\left\{(E_{11}-E_{22})^2+8\tau_1[\tau_1(E_{11}+E_{22})-\kappa H_{13}]\right\}^{1/2}\Big\}^{1/2}\ .
\end{eqnarray}
The character of the solutions will be either exponential or oscillatory depending on whether these quantities are real or imaginary respectively.
In constrast with the special cases $\kappa=0$ and $\tau_1=0$ previously considered, characterized by a fixed value of  the linear velocity, in the general case the dependence on $\nu$ must be taken into account. 

\subsection{The Pirani supplementary conditions}

The Pirani supplementary conditions (\ref{PcondFS}) require 
\beq
\label{Pconds}
S_{01}=0\ , \quad S_{02}=0\ , \quad S_{03}=0\ .
\eeq
Comparing the first two conditions with Eqs.~(\ref{s01sol}) and (\ref{s02sol}) we get 
\beq
h_\pm=l_\pm=0\ , 
\eeq
and 
\beq
\label{c0solP}
c_0=\kappa c_m\frac{\kappa^2+\tau_1^2-E_{22}}{2\kappa\tau_1-H_{13}}\ ,
\eeq
so that $S_{12}$ and the particle mass $m$ are both constant.
Eqs.~(\ref{s12sol}) and (\ref{masssol}) imply
\beq
\label{cmsolP}
c_m=\left[1-\frac{\kappa\tau_1}{\kappa\tau_1-H_{13}}\right]m\ .
\eeq
Next by substituting these values of the constants $c_0$ and $c_m$ into Eq.~(\ref{s12sol}), we obtain
\beq
\label{s12solP}
S_{12}=-\frac{m\kappa}{\kappa\tau_1-H_{13}}\ .
\eeq
Finally comparing the last of the Pirani conditions Eq.~(\ref{Pconds}) with  Eq.~(\ref{s03sol}) leads to 
\beq
c_\pm=d_\pm=0\ , 
\eeq
provided that $\omega_+\not=\omega_-$, implying that both the remaining components $S_{13}$ and $S_{23}$ are identically zero as well, from Eqs.~(\ref{s23sol}) and (\ref{s13sol}).
This case of constant solutions for the components of the spin tensor has been already considered previously \cite{bdfg2}.

\begin{figure} 
\typeout{*** EPS figure 1}
\begin{center}
\includegraphics[scale=0.4]{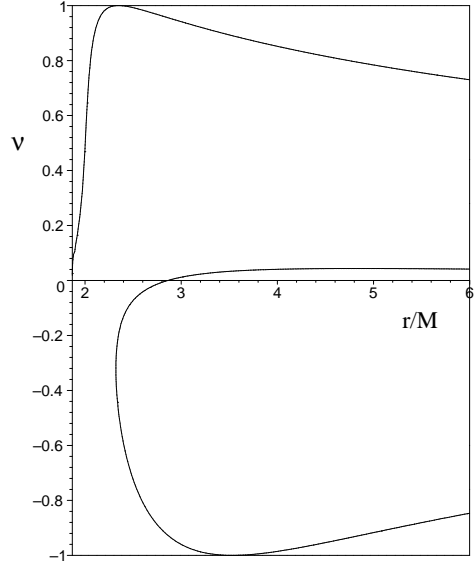}
\end{center}
\caption{The behaviour of the linear velocities satisfying the conditions (\ref{eqnuP}) is shown as a function of $r/M$ for $a/M=0.5$. The corresponding orbits become null at $r_{{(\rm geo)}+}/M\approx2.347$ and $r_{{(\rm geo)}-}/M\approx3.532$. The outer horizon is located at $r_+/M\approx1.866$.
}
\label{fig:1}
\end{figure}

Consider now the case $\omega_+=\omega_-\equiv\omega$, which implies 
\beq
\omega=\frac{\sqrt{2}}{2}[\kappa^2-\tau_1^2-E_{33}]^{1/2}\ ,
\eeq
together with the condition
\beq
\label{omeqcondP}
(\kappa^2-\tau_1^2)^2-[2(\kappa^2+\tau_1^2)-E_{33}]E_{33}-4\kappa\tau_1H_{13}=0\ .
\eeq
The last of (\ref{Pconds}) is satisfied also by setting $\omega^2=-\tau_1^2$ in Eq.~(\ref{s03sol}), implying $\kappa^2+\tau_1^2-E_{33}=0$; substituting then this relation in Eq.~(\ref{omeqcondP}) leads to
\beq
\label{omeqcondPnew}
\kappa\tau_1+H_{13}=0\ ,
\eeq
which gives the following equation for the linear velocity $\nu$
\begin{eqnarray}\fl
\label{eqnuP}
0&=&[k_{\rm (lie)}^2\nu_{\rm (gmp)}-H_{\hat r \hat \theta}]\nu^4
-[k_{\rm (lie)}^2(2\nu_{\rm (gmp)}^2+\nu_+\nu_-+1)+E_{\hat \theta \hat \theta}-E_{\hat r \hat r}]\nu^3\nonumber\\
\fl
&&+3k_{\rm (lie)}^2\nu_{\rm (gmp)}(1+\nu_+\nu_-)\nu^2
-\{k_{\rm (lie)}^2[2\nu_{\rm (gmp)}^2+\nu_+\nu_-(1+\nu_+\nu_-)]+E_{\hat r \hat r}-E_{\hat \theta \hat \theta}\}\nu\nonumber\\
\fl\
&&+k_{\rm (lie)}^2\nu_{\rm (gmp)}\nu_+\nu_- + H_{\hat r \hat \theta}\ .
\end{eqnarray}
For small values of the spacetime rotation parameter, the relevant solutions of the  above equation are given by
\beq\fl\quad
\nu \simeq\pm 2\left[\frac{M(r-9M/4)}{r(r-2M)}\right]^{1/2}-\frac38 a\frac{(1-2M/r)^{1/2}}{r^3}\frac{r-3M}{r-9M/4}(r^2+10Mr-27M^2)\ ,
\eeq
to first order in  $a$. As expected, in the limiting case of vanishing rotation parameter (Schwarzschild spacetime) the corresponding solutions reduce to 
\beq
\label{nupiranischw}
\nu =\pm 2\left[\frac{M(r-9M/4)}{r(r-2M)}\right]^{1/2}\ ,
\eeq
whose properties has been already discussed in \cite{bdfgj}.
Figure \ref{fig:1} shows the behaviour of the linear velocities satisfying the condition (\ref{eqnuP}) as functions of the radial coordinate for a fixed value of the ratio $a/M$.


Substituting the last of the Pirani conditions (\ref{Pconds}) into Eq.~(\ref{KerrFSeq3bKerrFSeq7b}) leads to
\beq
\label{s23solPomeq}
S_{23}= c \cos\tau_1\tau+d\sin\tau_1\tau\ ,
\eeq
where $\tau_1$ is evaluated at the allowed values of $\nu$ and $c$, $d$ are integration constants; Eq. (\ref{KerrFSeq2}) then gives 
\beq
\label{s13solPomeq}
S_{13}= c \sin\tau_1\tau-d\cos\tau_1\tau\ ;
\eeq
finally substituting condition (\ref{omeqcondPnew}) into Eq.~(\ref{s12solP}) leads to
\beq
\label{s12solPomeq}
S_{12}=-\frac{m}{2\tau_1}\ .
\eeq
The spin vector can thus be written as 
\beq
\pmatrix{
S^{1}(\tau)\cr
S^{2}(\tau)\cr
S^{3}(\tau)\cr}=
\pmatrix{\cos\tau_1\tau  &\sin\tau_1\tau & 0\cr
	-\sin\tau_1\tau  &\cos\tau_1\tau & 0\cr
         	0 & 0& 1\cr}
\pmatrix{
S^1(0)\cr
S^2(0)\cr
S^3(0)\cr}\ ,
\eeq
where 
\begin{eqnarray}
S^1(0)=c\ ,  \qquad
S^2(0)=d\ ,  \qquad 
S^3(0)=-\frac{m}{2\tau_1}\ . 
\end{eqnarray}
To first order in the rotation parameter, the first torsion turns out to be given by
\beq\fl\quad
\tau_1=\pm 2\frac{[M(r-9M/4)]^{1/2}}{r(r-3M)}-\frac38 a\frac{(r^2+2Mr-9M^2)^2}{r^3(r-9M/4)(r-3M)^2}+O\left(\frac{a^2}{M^2}\right)\ ,
\eeq
and in the weak field limit reduces to
\beq\fl
\tau_1=\pm 2\left(\frac{M}{r^3}\right)^{1/2}\left[1+\frac{15}{8}\frac{M}{r}+\frac{639}{128}\frac{M^2}{r^2}\right]-\frac38 \frac{a}{r^2}\left[1+\frac{49}{8}\frac{M}{r}\right]+O\left(\frac{M^4}{r^4},\frac{a^2}{M^2}\right)\ .
\eeq
It is worth noting that in the Schwarzschild case the spin vector does not precess with respect to a frame which is nonrotating at infinity (see \cite{bdfgj}), since $\tau_1=\Omega_U$, so that the angular dependence of the varying components $S^1$ and $S^2$ is just $\tau_1\tau=\phi$, using Eq.~(\ref{eq:phitau}). 
In the Kerr case instead $\tau_1\not=\Omega_U$ leading to $\tau_1\tau=\phi\,\tau_1/\Omega_U$, causing a net precession with respect to infinity.

Furthermore, it is easy to show that the spin vector is Fermi-Walker transported along $U$, i.e.
\beq\fl\quad
\label{fweq}
0=\frac{D_{(\rm fw)}S}{\rmd \tau_U}\equiv P(U)\frac{DS}{\rmd \tau_U}
=
 \left[\frac{\rmd S^{1}}{\rmd \tau_U}-\tau_1S^2 \right] E_1
+\left[\frac{\rmd S^{2}}{\rmd \tau_U}+\tau_1S^1 \right] E_2\ ,
\eeq
where $P(U)^\mu_\alpha=\delta^\mu_\alpha+U^\mu U_\alpha$  projects into the local rest space of $U$, by using Eqs. (\ref{KerrFSeq2}) and (\ref{KerrFSeq3}) together with the last of the Pirani conditions (\ref{Pconds}).

The spin invariant (\ref{sinv}) becomes in this case
\beq
s^2=c^2+d^2+\frac{m^2}{4\tau_1^2}\ .
\eeq
The Mathisson-Papapetrou model is valid  if the condition $|s|/(mM)\ll1$ is satisfied. 
From the previous equation it follows that the sum of the bracketed terms must be small, i.e. 
\beq
\label{scoeffs}
\frac{c^2}{m^2M^2}\ll1\ , \qquad \frac{d^2}{m^2M^2}\ll1\ , \qquad \frac1{4M^2\tau_1^2}\ll1\ .
\eeq
But these conditions cannot be satisfied for any allowed values of the radial coordinate, since the third term of (\ref{scoeffs}) is much smaller than $1$ only in the case of ultrarelativistic motion, which occurs only as $r\to r_{{(\rm geo)}\pm}$, where the orbits approach null geodesics (see Fig. \ref{fig:1}).

Finally, the total 4-momentum $P$ is given by 
\beq\fl
P = mU-\kappa S_{12} E_2 -\kappa S_{13} E_3
= \frac23c_mU+\frac{\kappa}{3\tau_1}c_mE_2-\kappa[c \sin\tau_1\tau-d\cos\tau_1\tau]E_3\ ,
\eeq
from Eqs.~(\ref{PdefFS}) and (\ref{Pscompts}).

\subsection{The Tulczyjew supplementary conditions}

It remains to discuss the case of Tulczyjew supplementary conditions (\ref{TcondFS}).
By solving for the first derivatives, a straightforward calculation shows that the set of equations (\ref{Tconds}) simplifies to
\begin{eqnarray}
\label{Tcond1}
\frac{\rmd S_{01}}{\rmd \tau_U}&=&-m\frac{S_{03}}{S_{13}}+\tau_1S_{02}\ , \\
\label{Tcond2}
\frac{\rmd S_{02}}{\rmd \tau_U}&=&-\tau_1S_{01}+\kappa S_{12}\ , \\
\label{Tcond3}
\frac{\rmd S_{03}}{\rmd \tau_U}&=&m\frac{S_{01}}{S_{13}}+\kappa S_{13}\ , \\
\label{Tcond4}
\quad
0&=&\frac{m}{S_{13}}[S_{01}S_{23}-S_{02}S_{13}+S_{03}S_{12}]\ ,
\end{eqnarray}
provided that $S_{13}\not=0$ is assumed.
Solving Eq.~(\ref{KerrFSeq1}) for $S_{02}$, substituting into Eq.~(\ref{Tcond2}) and using the first equation of (\ref{KerrFSeq5cKerrFSeq6c}) leads to
\beq
0=-\kappa\tau_1S_{01}+(\tau_1^2-E_{22})S_{12}+c_0\ .
\eeq
Substituting then the solutions (\ref{s12sol}) and (\ref{s01sol}) implies that both components $S_{01}$ and $S_{12}$ are constant, so that $S_{02}=0$ follows from Eq.~(\ref{s02sol}). Since $S_{03}$ must vanish according to Eq.~(\ref{Tcond1}), Eq.~(\ref{Tcond3}) together with the solution (\ref{s13sol}) require that $S_{13}$ vanishes as well. But this contradicts the assumption $S_{13}\not=0$, so only the case $S_{13}=0$ remains to be considered. 

If $S_{13}=0$  Eqs.~(\ref{TcondFS}) reduce to
\begin{eqnarray}
\label{Tcond1c2}
\frac{\rmd S_{01}}{\rmd \tau_U}&=&-m\frac{S_{02}}{S_{12}}+\tau_1S_{02}\ , \\
\label{Tcond2c2}
\frac{\rmd S_{02}}{\rmd \tau_U}&=&m\frac{S_{01}}{S_{12}}-\tau_1S_{01}+\kappa S_{12}\ , 
\end{eqnarray}
provided that $S_{12}\not=0$.
The requirement $S_{13}=0$ implies that the components $S_{03}$ and $S_{23}$ also vanish, as from Eqs. (\ref{s03sol}) and (\ref{s23sol}).
Solving Eq. (\ref{KerrFSeq1}) for $S_{02}$, substituting in Eq. (\ref{Tcond2c2}) and using the first equation of (\ref{KerrFSeq5cKerrFSeq6c}) leads to
\beq
\label{Tcond2c2b}
0=-\kappa c_m S_{01}+(c_0-E_{22}S_{12})S_{12}+(\kappa S_{01}-\tau_1S_{12})^2\ ;
\eeq
substituting then the solutions (\ref{s12sol}) and (\ref{s01sol}) implies that both components $S_{01}$ and $S_{12}$ are constant, so that $S_{02}=0$, from Eq. (\ref{s02sol}), and Eq. (\ref{Tcond1c2}) turns out to be identically satisfied.
The solutions for $S_{01}$ and $S_{12}$ can be obtained by substituting Eqs.~(\ref{s12sol}) and (\ref{s01sol}) into Eq.~(\ref{Tcond2c2}), setting $h_\pm=l_\pm=0$, solving for the integration constant $c_0$ and eliminating $c_m$ in favor of the mass $m$ through Eq.~(\ref{masssol})
\begin{eqnarray}
S_{01}&=&\frac{m}2\frac{(3\kappa\tau_1-H_{13})H_{13}-2\kappa^2E_{11}\mp(\kappa\tau_1-H_{13})\sqrt{\Psi}}{(\tau_1H_{13}-\kappa E_{11})(\tau_1^2-E_{11})}\ , \nonumber\\
S_{12}&=&\frac{m}2\frac{H_{13}\mp\sqrt{\Psi}}{\tau_1H_{13}-\kappa E_{11}}\ ,
\end{eqnarray}
where
\beq
\Psi=4\kappa^2E_{11}-(4\kappa\tau_1-H_{13})H_{13}\ ,
\eeq
which are in agreement with the condition $S_{12}\not=0$ assumed above. 
Hence it follows that $S_{01}=\nu_pS_{12}$, with
\beq
\nu_p=\frac12\frac{2\kappa\tau_1-H_{13}\pm\sqrt{\Psi}}{\tau_1^2-E_{11}}\ ,
\eeq
so that the spin vector must be constant and orthogonal to the plane of the orbit, having $S^2$ as its only nonvanishing component, and the total 4-momentum $P$ (see Eqs. (\ref{PdefFS}) and (\ref{Pscompts})) also lies in the cylinder of the circular orbit
\beq
P=mU-(\kappa+\nu_p\tau_1)S_{12}E_2\ .
\eeq
This solution, having constant spin components, was already found and discussed in a previous article \cite{bdfg2}.

\section{Conclusions}

Spinning test particles in  circular motion around a rotating Kerr black hole have been discussed in the framework of the Mathisson-Papapetrou approach supplemented by the usual ``intrinsic'' Pirani and Tulczyjew supplementary conditions, greatly facilitated by the use of the Frenet-Serret formalism. One sees that
the restriction to circular motion as well as the natural choice of Pirani conditions severely limit the solutions of the equations of motion,
as already shown by the corresponding analysis already done for the special case of the Schwarzschild black hole. Essentially the only interesting solutions are obtained by locking the spin vector precession to the Frenet-Serret rotational velocity of the path,  with a spin vector Fermi-Walker transported along an accelerated center of mass world line, but these violate
the test particle assumption except for orbits very close to the null geodesic case not far from the black hole horizon. In that case the spin vector must be aligned with the direction of motion from general considerations \cite{masspin2}.
The Tulczyjew conditions, instead, have no natural relationship to the Frenet-Serret properties of the particle path and do not admit such specialized solutions, instead producing only solutions where the spin vector is constant and orthogonal to the plane of the orbit.
Although these calculations seem rather academic, it is important to flush out the physical content of this only model for the motion of spinning test particles, in view of experiments like GP-B \cite{GPB} which are already in progress.

\end{document}